\documentclass[12pt]{iopart}
\usepackage{epsf,here,amssymb}
\begin{document}
\renewcommand{\theequation}{\thesection.\arabic{equation}}
\begin{titlepage} 
\vspace*{2cm}
\begin{center}
{\Large{Choice of the uncritical manifold for dilute polymer solutions}}\bigskip\bigskip\bigskip\bigskip\\
{\large Andrea Ostendorf} and {\large Johannes S.~Hager }\bigskip\\
Fachbereich Physik, Universit\"at-Duisburg-Essen, 45117 Essen, Germany \bigskip\\
\end{center}
\bigskip\bigskip\bigskip
\noindent
{\bf Abstract:} 
We discuss the dependence of the results of renormalized perturbation theory for dilute polymer solutions on the choice of the 
uncritical manifold where the perturbation series are evaluated. Special emphasis is given to the influence of polydispersity corrections on 
the results of one and two loop calculations. For monodisperse solutions we establish that after a Borel resummation the dependence 
on the choice of the uncritical manifold decreases when higher orders of the expansion are taken into account.
\bigskip\bigskip\\
email:\\
johannes@theo-phys.uni-essen.de\\
andrea@theo-phys.uni-essen.de
\bigskip\bigskip\\
It is a pleasure for us to dedicate this paper to Lothar Sch{\"a}fer on the occasion of his 60th birthday as a token of gratitude for his 
inspiration and advice.\\
\end{titlepage}

\setcounter{equation}{0}
\section{Introduction}
In now more than thirty years since de Gennes' observation \cite{dG72} that the correlations of a single long polymer chain can
 be mapped on the critical correlations of a $m$-component ferromagnetic spin model in the limit $m \rightarrow 0$, the
 application of dilatation symmetry to dilute polymer solutions has led to a rich and mature body of knowledge, which is
 able to explain the universal features found in experimental data. These developments have been reviewed in the book of
 des Cloizeaux and Jannink \cite{dC90} and with special emphasis on excluded volume effects in the monography of Sch{\"a}fer \cite{S00}. \\
Since exact results for realistic models are available only in space dimension $d=2$, quantitative calculations for the
physically interesting case $d=3$ heavily rely on the use of perturbation series expansions. Unfortunately, the expansion
parameter diverges in the critical limit of polymer length $n \rightarrow \infty$. This can be remedied by mapping the result
of perturbation theory on a uncritical manifold of the parameter space with the help of the renormalization group (RG). The RG
maps the physical variables chain length $n$, monomer size $l$ and excluded volume strength $\beta_e$ onto a set of renormalized
variables $(n_R,l_R,u)$, where low order perturbation theory for a given scaling function can be safely evaluated yielding
reasonable results. The determination of the uncritical manifold, by choosing the renormalized segment size $l_R$ introduces some
numerical parameters into the theory, which we will fix in the sequel by fitting our results for universal ratios to values
measured independently in simulations or experiments. Once those parameters are fixed, the theory has to prove its quantitative
accuracy by making predictions for additional measurable quantities. The necessity of such a fit procedure is clearly due to the
error which we introduce by calculating the perturbation expansion only to low order. With this approximation we break the 
strict scale invariance of the full renormalized scaling function under the RG map and introduce a dependence of the results on
the choice of the renormalized length scale $l_R$. We expect that the influence
of the choice of the uncritical manifold will vanish gradually when more and more orders of perturbation theory are
taken into account. One main goal of the present paper is to check this expectation for those few observables, namely the mean
square end-end-distance $R_e^2$, the radius of gyration $R_g^2$ and the second virial coefficient $A_2$, where the perturbation
series have been pushed to six, four and three loop order respectively. Since most of the more complex observables of interest
have been evaluated only at zero or one loop level, we spent some effort to describe the optimal choice of the renormalized manifold
at the one loop level as discussed in \cite{S00}. We also discuss how one can handle polydispersity effects within the theory, 
a important topic when we try to explain experimental measurements, which never work with purely monodisperse samples.
\\
 Our paper is organized as follows: In section 2 we define our polymer model and the observables of interest. We also review some
 basic notations of polydispersity. In section 3 we review the RG map and discuss the choice of the renormalized manifold at the
 one loop level. In section 4 we present the results of higher order perturbation theory and check their dependence on the choice
 of the uncritical manifold before and after a suitable Borel resummation. Section 5 gives a conclusion of our findings. A collection 
of perturbative results is presented in the appendix.
\setcounter{equation}{0}
\section{Polymer Model and Observable Quantities}
We represent a polymer chain of $n$ segments in a simple spring and bead model by $n+1$ beads, linearly connected to their 
neighbors by
elastic springs with mean distance $l$ and interacting with each other via a repulsive $\delta$-pseudo-potential of strength
$\beta_e$. This leads to the hamiltonian
\begin{equation}\label{1}
H\{\vec{r}_j\}={1\over4l^2} \sum_{j=1}^{n+1}(\vec{r}_j-\vec{r}_{j-1})^2+(4\pi l^2)^{d/2}\beta_e \sum_{j,j'}\delta(\vec{r}_j-\vec{r}_{j'}),
\end{equation}
where $\vec{r}_j$ is the position vector of the bead number $j$ and $d$ denotes the space dimension. The partition function
\begin{equation}\label{2}
Z=\int{{\cal D}[r] \; \mbox{e}^{-H\{\vec{r}_j\}}} \quad \mbox{with} \quad  {\cal D}[r]=\prod_{j=0}^n{d^dr_j\over(4\pi l^2)^{d/2}}
\end{equation}
for dimension $d>2$ can be calculated only in a perturbation expansion which orders in the parameter $z=\beta_e n^{2-d/2}$. Similar
considerations hold for averages of observables, defined as
\begin{equation}\label{3}
<O>={1\over Z} \int{{\cal D}[r]\; O\;\mbox{e}^{-H\{\vec{r}_j\}}} .
\end{equation}
In the sequel we focus our interest on the mean square end-end-distance
\begin{equation}\label{4}
R_e^2=<(\vec{r}_n-\vec{r}_0)^2>
\end{equation}
and the radius of gyration
\begin{equation}\label{5}
R_g^2=<{1\over n+1}\sum_{j=0}^n (\vec{r}_j-\vec{R}_{cm})^2> ,
\end{equation}
where $\vec{R}_{cm}={1\over n+1}\sum_{j=0}^n \vec{r}_j$ is the center of mass vector of the molecule.
The second virial coefficient can be read off from the virial expansion of either the osmotic pressure $\Pi$
\begin{equation}\label{6}
\Pi=c_p+{1\over2}A_2^{\Pi}c_p^2+O(c_p^3) ,
\end{equation}
where $c_p$ is the polymer concentration, or of the forward scattering intensity
\begin{equation}\label{7}
cI^{-1}(q=0,c_p,N)={1\over N_w}+ {A_2^S\over N_w}c+O(c^2),
\end{equation}
where $c=c_pN$ is the monomer concentration and $N_w$ is the weight averaged chain length defined as below. The two definitions
of $A_2^{\Pi}$ and $A_2^S$ coincide for monodisperse systems but differ for a general chain length distribution $P(n)$.
Both quantities can be obtained from the second virial coefficient $A_2(n_1,n_2)$ for two chains of lengths $n_1$ and $n_2$
according to \cite{S00} by averaging over the chain length distribution $P(n)$:
\begin{eqnarray}
A_2^{\Pi} &=& \sum_{n_1,n_2}P(n_1)P(n_2)A_2(n_1,n_2) \label{8} \\
A_2^S&=&\sum_{n_1,n_2}{n_1n_2\over N^2}P(n_1)P(n_2)A_2(n_1,n_2) \label{9} ,
\end{eqnarray}
where $N=\sum_n nP(n)$ is the  average chain length. Two other standard chain length averages that show up in the literature
are the weight average $N_w$ and the
$z$-average $N_z$ defined as
\begin{eqnarray}\label{10}
N_w:={1\over N}\sum_{n}P(n)n^2=N p_2 \quad \mbox{and} \quad N_z:={1\over N N_w}\sum_{n}P(n)n^3={p_3 \over p_2}N.
\nonumber\\
\end{eqnarray}
$N_w$ and $N_z$ can be expressed as indicated above in terms of the average chain length $N$ and the second and third 
moments $p_2$ and $p_3$ of the reduced chain length distribution $p(y)$ defined by
\begin{equation}\label{11}
P(n)={1\over N}p({n \over N}).
\end{equation}
\setcounter{equation}{0}

\section{Renormalization}
In order to map our perturbative results to the uncritical manifold we introduce renormalized variables according to 
\begin{eqnarray}
l&=&\lambda  l_{R}\label{12} \\
n&=&\lambda^{-2}n_R Z_n(u) \label{13} \\
\beta_e&=&\lambda^{\epsilon}u Z_u(u)  \label{14} ,
\end{eqnarray}
where $\epsilon=4-d$ and the scaling parameter $\lambda\in [0,1]$ measures the degree of dilatation. For finite polymer
concentration $c_p$ and finite momentum $q$ we define renormalized quantities via $c_{pR}=c_pl_R^3$ and $q_R=q l_R$. The
$Z$-factors $Z_n={Z_2\over Z_{\phi}}$ and $Z_u={Z_4\over Z_{\phi}}$ have been calculated for $\phi^4$ field theory in the
minimal subtraction scheme to five loop order (see \cite{VSF91} and refs. in there) and are given in the appendix for $\epsilon=1$. 
The dependence of the $Z$-factors
on a change of the scaling parameter $\lambda$ can be obtained via integration from the flow equations
 \begin{eqnarray}
\lambda{d u\over d \lambda}&=&W(u_{R})\label{15} \\
\lambda{d \over d \lambda}\ln({Z_2\over Z_{\phi}})&=&  2-{1\over \nu(u)} \label{16}\\
\lambda{d \over d \lambda}\ln(Z_{\phi})&=& \eta(u) ,\label{17}
\end{eqnarray}
where the Wilson function $W(u)$, except for a linear dependence on $\epsilon$, and the exponent functions $\eta$ and $\nu$
depend only on the renormalized coupling $u$. Since the perturbation series for $W,\eta$, and $\nu$ are only asymptotic, they
have to be resummed to yield reliable results. We follow the work of Schloms and Dohm \cite{SD89} who resummed the flow functions
at the upper critical dimension $d_c=4$ and then evaluated the flow equations and renormalized scaling functions directly in $d=3$
dimensions, without further expansion in $\epsilon$. Besides the Gaussian fixed point  at $u=0$ the Wilson function $W(u)$ 
has a nontrivial fixed point at $u^*=0.364$, which is related to the excluded volume limit $n \rightarrow \infty, \beta_e>0$.
The correlation length exponent $\nu$, which governs the power law $R^2\sim N^{2\nu}$ for $R_g^2$ and $R_e^2$ in the excluded
volume limit, takes the fixed point value $\nu(u)=0.588$. To measure the distance from the excluded volume fixed point, we introduce
the parameter $f={u \over u^*}$. We can now use our perturbative results for the observables $R_e,R_g$ and $A_2^S$ to form
universal ratios, i.e. combinations depending only on $f$ and global characteristics of the system like space dimension or
polydispersity. Such quantities reduce to pure numbers at the fixed points of the RG. Just like the critical exponents, they are
universal in the sense that they are independent of the microstructure of the underlying model. From our three observables we can
form two independent ratios
\begin{equation}\label{18}
R^2_{g/e} = {6R^{2}_g \over R^2_e} \quad \mbox{and}\quad \psi^S=\left({d \over 12 \pi}\right)^{d/2} {A_2^S \over R_g^d},
\end{equation}
where the prefactor of the interpenetration ratio $\psi^S$, which roughly measures the volume that a chain excludes for other
chains, has purely historical reasons. For polydisperse systems, where $A_2^S$ and
$A_2^{\Pi}$ differ, another ratio can be introduced by using $A_2^{\Pi}$.

\subsection*{Choice of the uncritical manifold in one loop approximation}

As a general recipe, the renormalized length scale $l_R$ should be chosen smaller than the smallest macroscopic length scale
important  for the observable of interest. For example, it does not make sense to choose $l_R>R_g$, because then the whole coil
would be smaller than one effective segment of size $l_R$. In lowest order perturbation theory we find $R_g^2={p_3\over p_2}l^2N$,
which after renormalization gives
\begin{equation} \label{19}
R_g^2=l_R^2N_{zR} .
\end{equation}
Thus the choice $N_{zR}\approx 1$ fixes $l_R$ as $l_R\approx R_g$. Note that with this procedure we introduce some polydispersity
dependence into the choice of the uncritical manifold. We will discuss this later in the section. Now finite polymer concentration
$c_p$ and finite momentum $q$ both introduce additional characteristic length scales into the system \cite{S00} and depending on
their values and the observable of interest it can be necessary to choose $l_R^d\approx {1\over 2fc_p}$ or $l_R \approx {1\over q}$ in the
appropriate limits. Following \cite{S00} we choose the relation
\begin{equation} \label{rp}
{\bar{q}^2 \over q_0^2}+{n_0 \over N_{zR}}+f{c_R \over c_0}=1 .
\end{equation}
to interpolate smoothly between the above limits. The constants $q_0^2,n_0$ and $c_0$ have been introduced in (\ref{rp}), since our
qualitative arguments fix $l_R$ only up to a constant and the dependence of our results on the parameters $q_0^2,n_0$ and $c_0$
displays the approximation we made by truncating the perturbation expansion at low order. In the sequel we will concentrate on
the dilute low momentum limit $c_p\rightarrow 0,q \rightarrow 0$, where (\ref{rp}) reduces to $N_{zR}=n_0$. For readers interested
in the determination of $q_0^2$ and $c_0$ we refer to \cite{S00}.
Following  \cite{S00} we use the fixed point value of the interpenetration ratio 
$\Psi^*=\Psi^{(S)}(u^*)$ to fix $n_0$, since it depends on $N_{zR}$ already in the zero loop 
approximation.

\begin{figure}[H]
\epsfxsize 10.5cm
\hspace{1cm}
\epsffile{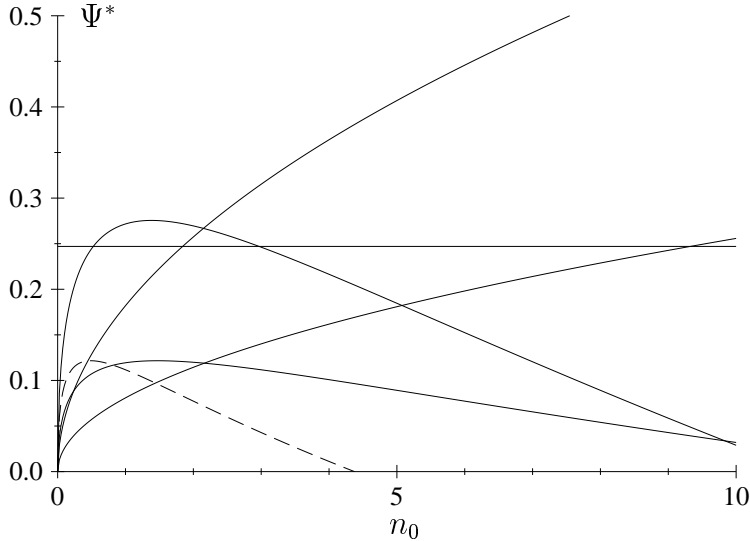}
\caption{Zero and one loop results for the universal ratio  $\Psi^*$ plotted versus $n_0$:  The upper curves display the result 
 for a monodisperse ensemble and the lower ones for a exponential ensemble. The dashed line shows the one loop result for an 
exponential ensemble evaluated with the choice $N_R=n_0$.}\label{fig1}
\end{figure}

For a monodisperse ensemble $(p_2=1)$ and an exponential ensemble $(p_2=2)$ the zero and one loop approximations 
are displayed in figure \ref{fig1} 
together with the most precise value $\Psi^*=0.247$ obtained from monodisperse computer simulations 
\cite{Nic91} which is in fair agreement with the best experimental value $\Psi^*=0.245\pm 0.005$ 
\cite{C80}. While the zero loop result reproduces the simulation result for a value $n_0=1.85$
the one loop result allows the choices $n_0=0.53$ and $n_0=2.95$ to fit the simulation data. 
In \cite{S00} the value $n_0=0.53$  was chosen mainly because of the unreasonably large polydispersity corrections that occur in
$\Psi^*$, when the relation $N_R=n_0$ is used for the determination of the uncritical manifold. We can resolve this problem by
absorbing the polydispersity dependence partially in the choice $N_{zR}=n_0$ (corresponding to $l_R \approx R_g$) of the
uncritical manifold. For the one loop result of the exponential ensemble both
choices of the uncritical manifold are included in figure |ref{fig1}.  
One finds that our choice $N_{zR}=n_0$ allows us to choose  $n_0=2.95$ with a
value for $\Psi^*_e=0.112$ consistent with the zero loop result  $0.11$ for $n_o=1.85$ and  close to the value $0.12$ found with 
$N_R=n_0=0.53$ in  \cite{S00}.
Unfortunately presently neither experimental nor simulation data on the polydispersity dependence of  $\Psi$ are available. We prefer the
choice $n_0=2.95$ since it enhances the numerical precision of several one loop results for intra chain properties.
As an example we consider two universal ratios $\sigma_{R}$ and $\delta$ which have been studied in the literature on di-block
copolymers. We divide the chain into two blocks of relative length $x_1={n_1 \over n}$ and $x_2={n_2 \over n}$, which may have
different chemical compostition.  This setup allows for two different values $u_{11}$ and $u_{22}$ of the intra-block excluded
volume repulsion and for a third value $u_{12}$ of the inter-block excluded volume repulsion. The ratios  $\sigma_{R}$ and
$\delta$ are defined as
\begin{equation}\label{21}
\sigma_{R} = {R^2_e \over R^2_{e1H}+R^2_{e2H}} \quad \mbox{and}\quad \delta={R^2_g-x_1 R^2_{g1H}-x_2 R^2_{g2H} \over
2x_1x_2(R^2_{g1H}+R^2_{g2H})},
\end{equation}
where the subscript $H$ denotes zero inter-block coupling $u_{12}=0$. At the symmetrical fixed point $u_{12}=u_{11}=u_{22}=u^*$ the
ratios  $\sigma_{R}$ and $\delta$ can be evaluated explicitly. Using the asymptotic power law $R \sim n^{\nu}$ we find \cite{MFC89}
\begin{equation}\label{22}
\sigma^*_{R} = {1 \over x_1^{2\nu}+(1-x_1)^{2\nu}} \quad \mbox{and}\quad \delta^*={1-x_1 ^{2\nu}-(1-x_1)^{2\nu} \over
2x_1(1-x_1)(x_1^{2\nu}+(1-x_1)^{2\nu}}.
\end{equation}
For other values of the excluded volume strength we can evaluate $\sigma_{R}$ and $\delta$ only perturbatively and the result
again depends on the choice of the renormalized manifold. The radii of gyration of both blocks are additional length scales which
have to be considered in the choice of the renormalized manifold. In the definition of  $\sigma_{R}$ and $\delta$ one observes
that in the limit $x_i \rightarrow 0$ the contribution of the smaller block $x_i$ can be neglected compared to the contribution of
the larger block. Thus we can safely use $N_{1zR}+N_{2zR}=n_0$ for the determination of the uncritical manifold, as in the
homo-polymer case. 

\begin{figure}[H]
\epsfxsize 10.5cm 
\hspace{1cm}
\epsffile{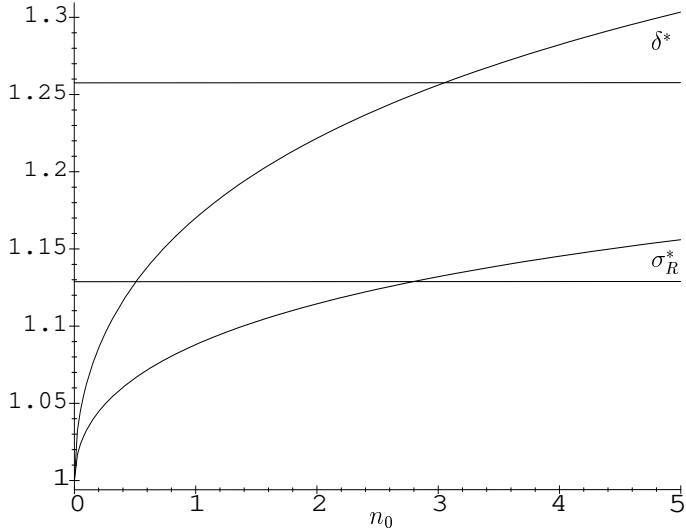}
\caption{One loop and exact results for the universal ratios  $\sigma_{R}^*$ and $\delta^*$ for equal block size $x_1=x_2={1 \over 2} $ 
plotted versus $n_0$ }\label{fig2}
\end{figure}

\begin{figure}[H]
\epsfxsize 10.5cm 
\hspace{1cm}
\epsffile{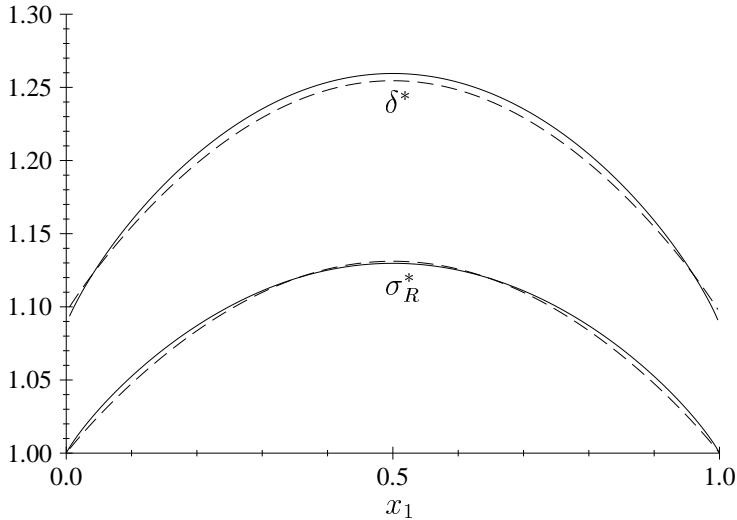}
\caption{One loop (dashed) and exact (full lines) results for the universal ratios  $\sigma_{R}^*$ and $\delta^*$ for $n_0=2.95$ 
plotted versus $x_1$}\label{fig3}
\end{figure}

 Figure \ref{fig2} displays the $n_0$ dependence of the one loop results for $\sigma_{R}$ and $\delta$ for the
symmetric case $x_1=x_2={1\over 2}$, together with the exact values of eq. (\ref{22}). One finds that the choice $n_0=2.95$
reproduces the exact values within a accuracy of $2\%$. In figure \ref{fig3} we plotted the exact and one loop results for the fixed point
values $\sigma_{R}^*$ and $\delta^*$ as a function of the relative block size $x_1$. The perturbative results deviate from the
exact ones (using $\nu=0.588$ from high order perturbation theory) by less than $2\%$ over the whole interval of block compostitions.
Similar results have been found for a one loop calculation of the persistence length $L_p$ \cite{ET}. 
\\
The limitations of the one loop approximation can be judged from the result for the $n_0$ dependence of the universal ratio $R^{2}_{g/e}$
displayed in figure \ref{fig5}. For $n_0=2.95$ the value $R^{2*}_{g/e}=0.983$ is still closer to the value $1$ for noninteracting chains than to the
value $0.96$ obtained in high precision simulations of self-avoiding chains at the excluded volume fixed point \cite{LMS95}.

\setcounter{equation}{0}
\section{Higher orders of perturbation theory}
\begin{figure}[H]
\epsfxsize 10.5cm 
\hspace{1cm}
\epsffile{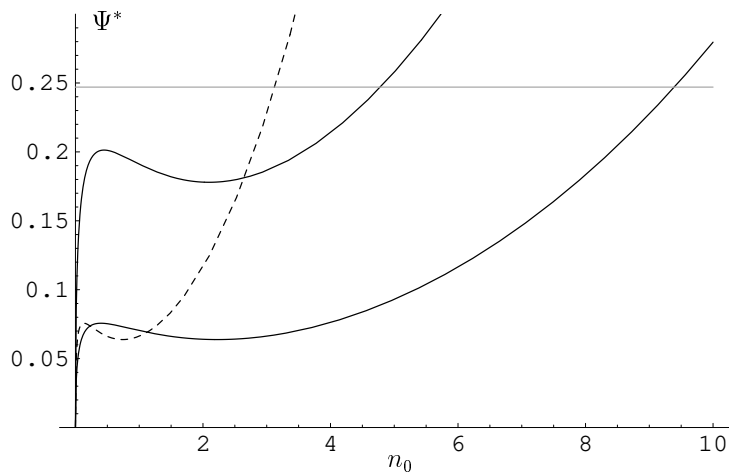}
\caption{Two loop results for $\Psi^*$ plotted versus $n_0$. The upper full curve corresponds to a monodisperse ensemble. 
The lower full and dashed curves correspond to a exponential ensemble evaluated  with $N_{zR}=n_0$ and $N_R=n_0$ 
respectively.}\label{fig4}
\end{figure}

\begin{figure}[H]
\epsfxsize 10.5cm 
\hspace{1cm}
\epsffile{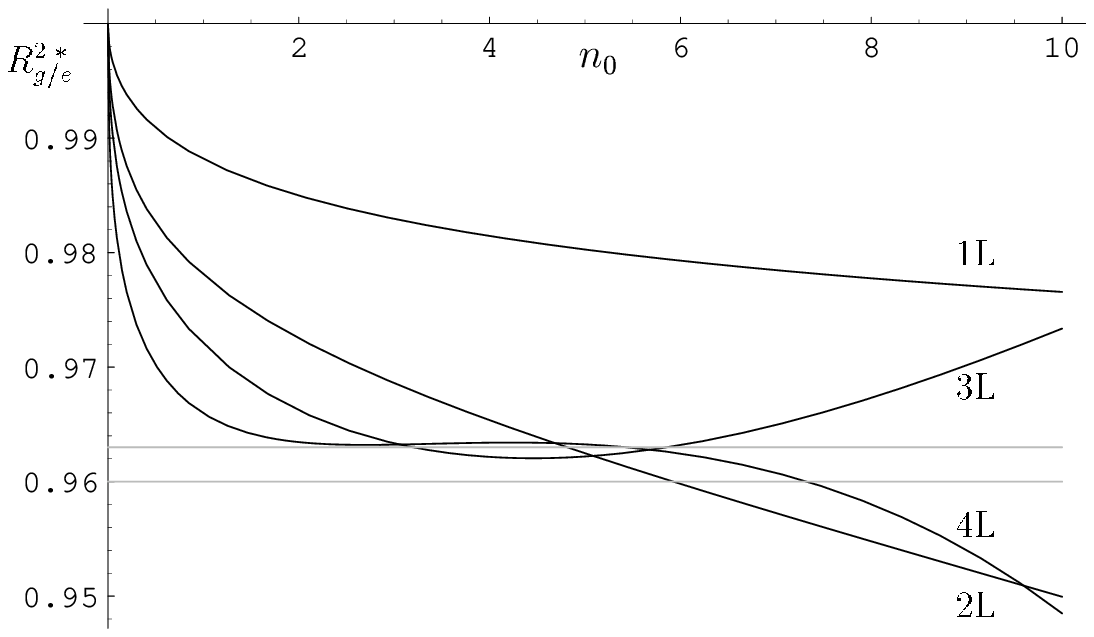}
\caption{One to four loop results for the universal ratio $R^{2*}_{g/e}$ plotted versus $n_0$.}\label{fig5}
\end{figure}

For only a few polymer observables, namely $R_g, R_e$ and $A_2$ the loop expansion has been pushed beyond the two loop level.
Nickel and coworkers calculated $R_e$ up to the order of six loops \cite{MN84,MN87} and $R_g$ to four loops \cite{SN90} and $A_2$ to two 
loops for a monodisperse ensemble. Their results
enable us to investigate the $n_0$ dependence of the universal ration $R^2_{g/e}$ up to the four loop level. In order to further
study the behavior of $\Psi^S$ we calculated the three loop contribution to the second virial coefficient $A_2(n,n)$ \cite{SOH99}, which allows
the evaluation of $\Psi^S$ at the three loop level for a monodisperse ensemble. In addition we calculated the two loop contribution
to $R_g^2$ and $\Psi$ for arbitrary polydispersity. \\
First we want to test wether a consistent choice of $n_0$ is possible at the two loop level. Inspecting figure \ref{fig4} we find that the two loop 
result
for $\Psi^*$ reproduces the value $\Psi^*=0.247$ for a value of $n_0=4.78$. Furthermore one observes that the $n_0$ dependence of $\Psi^*$ 
is rather weak for a range of values $n_0\in [0.5,4]$ but at a value of $\Psi^*\approx 0.18-0.2$ well below the desired result. Using the condition 
$N_{zR}=n_0$ leads to $\Psi^*_e=0.085$ for the exponential ensemble, somewhat below the zero and one loop results. The
choice $N_{R}=n_0$ (dashed curve) instead does not allow for a consistent  fit with reasonable polydispersity corrections. With the choice
$n_0=4.78$ we can read off the prediction $R^{2*}_{g/e}=0.963$ from the two loop result in figure \ref{fig5}, a value that compares already well 
with the Monte-Carlo result $0.96$ and with the two loop epsilon expansion result $0.959$ \cite{BM85}. \\

\begin{figure}[H]
\epsfxsize 10.5cm 
\hspace{1cm}
\epsffile{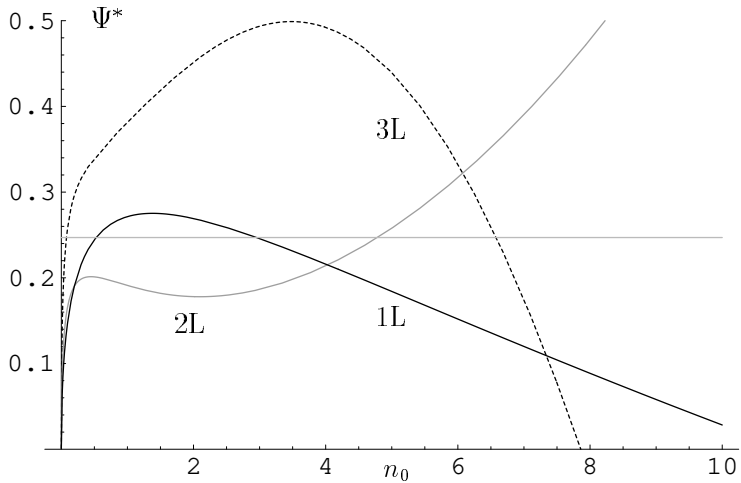}
\caption{One to three loop results for the universal ratio $\Psi^*$ plotted versus $n_0$.}\label{fig6}
\end{figure}

Beyond the two loop level the three and four loop results for $R^{2*}_{g/e}$ displayed in figure \ref{fig5} show the formation of a plateau 
at the value 
$0.963$ for  $n_0\in [2,6]$, nursing the hope that the choice of $n_0$ becomes less important with increasing order of perturbation theory. 
On the other hand our three loop result for $\Psi^*$ in figure \ref{fig6} exhibits a pronounced  $n_0$ dependence that seems to contradict our expectation. We can trace back this behavior to the asymptotic nature of the renormalized perturbation series, which have to be resummed in 
order to extract sensible information beyond the two loop level.

\subsection*{Resummation}

It is well known that the perturbation expansions in quantum field theory usually are only asymptotic series with zero radius of convergence 
\cite{KS80}. This gives rise to an exponential growth of the expansion coefficients. The leading behavior of the coefficients in high order  
perturbation theory can be obtained from a semiclassical calculation  as \cite{L77,BGZ77}
\begin{equation}\label{99}
\beta_{k}=k! \,(-a)^{k}\, k^{b}\, c\,\left(1+{\cal O}({1 \over k})\right) ,
\end{equation}
where in $d=4$ the coefficients are  $a={3\over 2}$ for our definition of the coupling and $b=2+M$ for a correlation function 
involving $M$ polymer chains.
With the knowledge of the asymptotic behavior we perform a standard Borel resummation procedure of our perturbation series for 
$R_e^2, R_g^2$ and $A_2^S$ as described in \cite{KS80,ZJ81}. The coefficients $b_0 \geq b+{3\over 2}$ and $\alpha$ involved 
in the resummation procedure have been tuned to $b_0=6$ and $\alpha=1$ in order to give optimal convergence of the approximations.
Note that despite the fact that we evaluate the perturbation series directly in $d=3$ dimensions we were forced to use the $d=4$ result 
$a={3\over 2}$ in order to obtain good convergence of the resummed series. This may be traced back to the fact that for the renormalization we used $Z$-factors which have been defined via minimal subtraction of $\epsilon$ poles at dimension $d=4 (\epsilon=0)$. These $Z$-factors 
also are given as asymptotic series and the stronger exponential growth of their coefficients seems to dominate the 
asymptotics of the renormalized series at
 $d=3$. A similar procedure was used in the calculation of universal quantities of the $O(m)$ symmetric $\phi^4$ model \cite{SD89}.

\begin{figure}[H]
\epsfxsize 10.5cm 
\hspace{1cm}
\epsffile{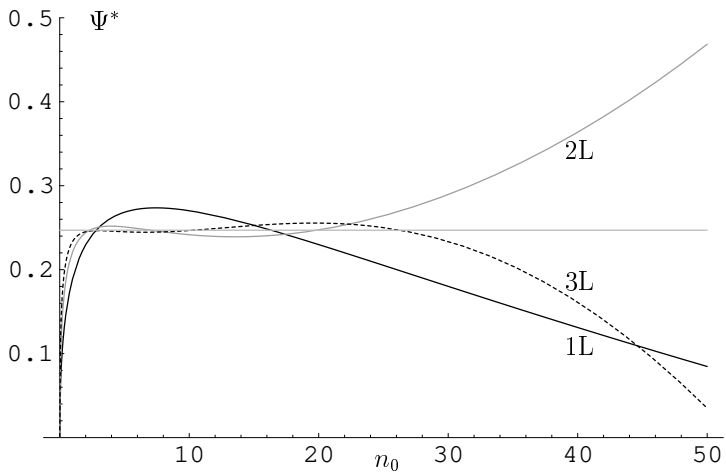}
\caption{Resummed one to three loop results for the universal ratio $\Psi^*$ plotted versus $n_0$.}\label{fig7}
\end{figure}

\begin{figure}[H]
\epsfxsize 10.5cm 
\hspace{1.0cm}
\epsffile{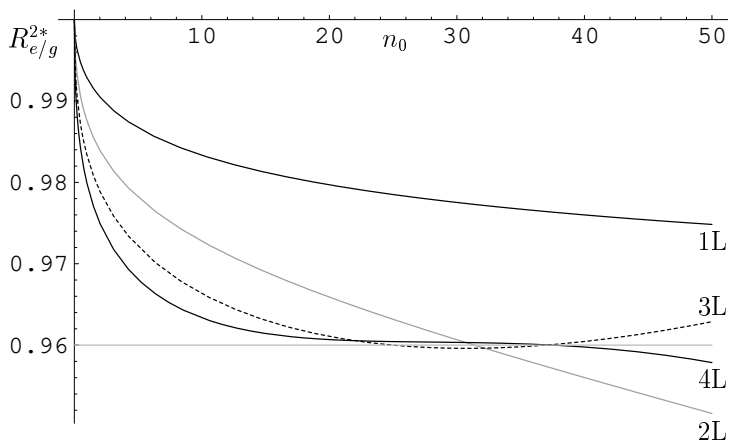}
\caption{Resummed one to four loop results for the universal ratio $R^{2*}_{g/e}$ plotted versus $n_0$.}\label{fig8}
\end{figure}

Figures \ref{fig7} and \ref{fig8} display the resummed results for $\Psi^*$ and $R_{g/e}^{2*}$ obtained via (\ref{18}) from the resummed 
functions for 
$R_e^2, R_g^2$ and $A_2^S$. For both ratios the variation as a function of $n_0$ is greatly reduced - the range of $n_0$ in figures 
\ref{fig7} and \ref{fig8} is extended by a factor $5$ compared to figures \ref{fig5} and \ref{fig6}. 
The most prominent effects can be seen in the two  and three loop results for $\Psi$.
The plateau region, where the two loop result was insensitive to $n_0$ is shifted close to the value expected from simulations. The variation of  
the three loop result is greatly reduced, being now fairly insensible to $n_0$ in the interval $n_0\in[2,30]$ for a value around $0.247$.  The 
effects of the resummation on $R_{g/e}^{2*}$ are  less dramatic. Mainly the value of the plateau already present in figure \ref{fig5} is shifted from 
$0.963$ to $0.96$ and thus is now in full accord with the simulation results.  

\section{Conclusions}
We studied  the dependence of several universal ratios on the choice of the uncritical manifold, where the evaluation of renormalized perturbation 
theory  gives sensible results. We found that the inclusion of the polydispersity dependence of the zero loop result for the radius of gyration into  the 
choice of the uncritical manifold allows for the choice $n_0=2.95$ leading to reasonable polydispersity corrections and to improved one loop 
estimates for several universal ratios. Furthermore we find this procedure necessary in order to obtain a consistent choice of $n_0$ at the two loop 
level. It would be interesting to check our predictions on the polydispersity dependence of $\Psi^*$ by comparison with precise experimental or 
simulation data. Beyond the two loop approximation we found that a resummation of the asymptotic series is mandatory to establish the increasing 
insensitivity of the results on the choice of $n_0$, which was expected on theoretical grounds. The resummed series for $\Psi^*$ and $R^{2*}_{g/e}$ 
display extended plateau regions where the $n_0$ dependence is weak, at values $\Psi^*=0.247$ and $R^{2*}_{g/e}=0.96$ in full agreement with
experimental measurements and Monte-Carlo simulations.

\begin{appendix}
\setcounter{equation}{0}
\section{Perturbative results}
\subsection*{Bare monodisperse results}
The perturbation series for $R_e^2, R_g^2$ and $A_2(n,n)$ evaluated directly in $d=3$ dimensions are
\begin{eqnarray}
R_e^2 &=& 6 R_0^2(1+{4 \over 3}z+({28\pi \over 27}-{16 \over 3})z^2+6.2968797z^3-25.0572507z^4\nonumber 
\\
&&+116.134785z^5-594.71663z^6),\label{a1}
\\
R_g &=&R_0^2(1+{134 \over 105}z+({1247\pi \over 1296}-{536 \over 105})z^2+6.564897z^3-26.70629z^4),\label{a2}
\nonumber \\ \\
A_2 &=& (4\pi)^{3/2}R_0^3 z(1-{32(7-4\sqrt{2}) \over 15}z+13.92783z^2-80.30z^3),\label{a3}
\end{eqnarray}
where $R_0^2=nl^2$ and $z=\beta_en^{1/2}$.

\subsection*{$Z$-factors}
The $Z$-factors as obtained from $\phi^4$ field theory \cite{VSF91} evaluated for $d=3$ are
\begin{eqnarray}
Z_n &=& 1-u-{7 \over 8}u^2-1.2708333u^3-5.299419u^4+40.504065u^5,\label{a4}
\\
Z_u&=&{1\over 2}(1+4u+{43\over 4}u^2+43.639293u^3+7.439240u^4),\label{a5}
\\
Z_n^{{1\over 2}} &=&1-{1 \over 2}u-{9 \over 16}u^2-0.91666667u^3-3.266246159u^4,\label{a6}
\end{eqnarray}

\subsection*{Copolymer quantities}

The $Z$-factors for the copolymer case at one loop order evaluated for $d=3$ are \cite{S00,SK93}
\begin{eqnarray}
Z_u^{(aa')}(u_{aa'})&=&{1 \over 2}\left( 1+(u_{aa}+u_{a'a'})+2 u_{aa'} +{\cal O}(u^2)\right) \label{aco8} \\
Z_N^{(a)}(u_{aa})&=&1-u_{aa} +{\cal O}(u^2) . \label{aco9}
\end{eqnarray}
The direct evaluation of renormalized perturbation theory in $d=3$ dimensions leads to
\begin{eqnarray}
R^2_{ga}&=&l_R^2n_{aR}\Bigg( 1+u_{aa} \left( {67 \over 105}n_{aR}^{1/2}-1 \right) \label{aco14} + {u_{a\bar{a}} n_{aR}^{1/2} \over 105} 
 \left( 384\kappa_{aR}^{7/2}+448\kappa_{aR}^{5/2} \right. \nonumber\\
& &\left. +13-{384\kappa_{aR}^4+640\kappa_{aR}^3+176\kappa_{aR}^2-32\kappa_{aR}+13 \over (1+\kappa_{aR})^{1/2}} \right)\Bigg) 
\end{eqnarray}
for the radius of gyration of block $a$, where $\kappa_{aR}={n_{\bar{a}R} \over n_a}$ denotes the renormalized ratio of the length of both blocks and to
\begin{eqnarray}\label{aco15}
R^2_{g}\!&=&\! l_R^2 (n_{1R} +n_{2R}) \nonumber\\
&&\Bigg( 1+{u_{11} \over 105(1+\kappa_{1R})^3}\left( (67+196\kappa_{1R})n_{1R}^{1/2}-(105+315\kappa_{1R}) \right) \nonumber \\
&&+ {u_{22} \over 105(1+\kappa_{2R})^3}\left( (67+196\kappa_{2R})n_{2R}^{(1/2)}-(105+315\kappa_{2R}) \right) \nonumber \\
&&+ {u_{12} (n_{1R}+n_{2R})^{1/2} \over 105(1+\kappa_{1R})^4} \bigg( 67(1+\kappa_{1R}^4) -67(1+\kappa_{1R})^{1/2}(1+\kappa_{1R}^{7/2}) \nonumber \\
& & -196(1+\kappa_{1R})^{1/2}(\kappa_{1R}+\kappa_{1R}^{5/2}) +268(\kappa_{1R}+\kappa_{1R}^{3})+402\kappa_{1R}^2\bigg)\Bigg) \nonumber \\ 
\end{eqnarray}
for the radius of gyration of the whole chain.
For the renormalized mean squared end-end distances we find
\begin{eqnarray}\label{aco20}
R^2_{ea}\!\!&=&\!\!6l_R^2n_{aR} \nonumber\\
&&\!\left( 1+u_{aa}\!\left( {2 \over 3}n_{aR}^{1/2}-1 \right)\!+\!{2u_{a\bar{a}} \over 9}n_{aR}^{1/2}\left( 1-8\kappa_{aR}^{3/2}+{8\kappa_{aR}^{2}+4\kappa_{aR}-1 \over (1+\kappa_{aR})^{1/2}} \right) \right) \nonumber \\ \\
R^2_{e}&=& 6l_R^2 (n_{1R} +n_{2R})\Bigg(1+u_{11}{{2 \over 3}n_{1R}^{1/2}-1 \over 1+\kappa_{1R}} +u_{22}{{2 \over 3}n_{2R}^{1/2}-1 \over 1+\kappa_{2R}} \nonumber \\
&&+{2u_{12} \over 3}(n_{1R}+n_{2R})^{1/2}\left(1-{1+\kappa_{aR}^{3/2} \over (1+\kappa_{aR})^{3/2}}\right)\Bigg) . \label{aco21}
\end{eqnarray}

\subsection*{Renormalized polydisperse results}
For a general chain length distribution $p(y)$ we obtain the following two loop result for the second virial coefficients $A^{\pi}_2$ and $A^S_2$, 
where $m=0$ corresponds to $A^{\pi}_2$ and $m=1$ to $A^S_2$. The polydispersity correction $c_{A}^m$ vanishes for a monodisperse ensemble.
\begin{eqnarray}\label{polyapp.1}
(4\pi l_{R})^{-\frac{3}{2}} A_{2}^m
\tilde{p}_{m+1}^{-2}&=&\bar{a}_{2}(v,c_{A}^m) \nonumber \\
&=&\frac{u^{\ast}}{2} f N_{R}^2
\bigg(1+u^{\ast}f\left(2-\sqrt{N_{R}} \frac{16}{15}\left(7-4\sqrt{2}\right)\right)+u^{\ast 2} f^2  \nonumber \\
&&\left(2- \frac{88}{15}
\sqrt{N_{R}}(7-4\sqrt{2})+\frac{N_{R}}{4}\left(\frac{1622}{15}-\frac{131
  \pi}{12} \right.\right.\nonumber \\
& &\left.\left.-\frac{1024 \sqrt{2}}{15}+\frac{32 \pi}{3} \ln2 +
\frac{125}{6} \arctan\frac{3}{4}\right)\right) + c_{A}^m \bigg) ,\nonumber \\
\end{eqnarray}
\begin{eqnarray}\label{polyapp.2} 
c_{A}^m&=& -\frac{16}{15}\sqrt{N_{R}} u^{\ast}f \: \bigg[ \frac{5}{\tilde{p}_{m+1}}
\int^\infty_0 \!\! dy \;  y^{m+\frac{3}{2}}\:p(y)+ 
\frac{2\:\tilde{p}_{m}}{\tilde{p}_{m+1}^{2}} 
\int^\infty_0\!\!dy\;y^{m+\frac{5}{2}}\:p(y) \nonumber \\
& & -\frac{1}{\tilde{p}_{m+1}^2 } 
\int^\infty_0 \!\! dy_{1} \int^\infty_0 \!\!
dy_{2} \; y_{1}^m y_{2}^m (y_{1}+y_{2})^\frac{5}{2} 
\:p(y_{1}) p(y_{2})-7+4 \sqrt{2} \bigg] \nonumber \\
&&+u^{\ast 2} f^2 \: \bigg[ \:2\: \bigg(   \frac{1} { \tilde{p}_{m+1}^2 }-1 \bigg)-\frac{88}{15}
\sqrt{N_{R}}    \bigg(\:\frac{5}{\tilde{p}_{m+1}}
\int^\infty_0 \!\! dy\; y^{m+\frac{3}{2}}\:p(y) \nonumber \\ &&+\frac{2\:\tilde{p}_{m}}{\tilde{p}_{m+1}^{2}}
\int^\infty_0\!\!dy\;y^{m+\frac{5}{2}}\:p(y) \nonumber \\ &&
 -\frac{1}{\tilde{p}_{m+1}^2 } 
\int^\infty_0 \!\! dy_{1} \int^\infty_0 \!\!
dy_{2} \; y_{1}^m y_{2}^m (y_{1}+y_{2})^\frac{5}{2} 
\:p(y_{1}) p(y_{2})-7+4 \sqrt{2} \bigg)+ \nonumber\\&&
+ \frac{N_{R}}{4} \: \bigg\{\frac{64}{3    \tilde{p}_{m+1}^2} \: \left(\int^\infty_0
\!\!dy \; y^{m+\frac{3}{2}} \:p(y)\right)^2 + \left(\frac{128}{3}-\frac{17 \:
  \pi}{3}\right)    \frac{\tilde{p}_{m+2}}{\tilde{p}_{m+1}} \nonumber\\
&& +\frac{406}{15    \tilde{p}_{m+1}^{  2}}    \int^\infty_0 \!\! dy_{1} \;
y_{1}^{m+\frac{5}{2}}   p(y_{1}) \int^\infty_0 \!\! dy_{2} \;
y_{2}^{m+\frac{1}{2}}   p(y_{2})\nonumber \\&&
+ \left(\frac{256}{15}-\frac{63  \pi}{12}\right) \frac{\tilde{p}_{m+3} 
  \tilde{p}_{m}}{\tilde{p}_{m+1}^{  2}}  
 - \frac{256}{15    \tilde{p}_{m+1}^{  2}} \int^\infty_0 \!\! dy_{1}
   \int^\infty_0 \!\! dy_{2}   
 \big(\:2\:y_{1}^{m+\frac{3}{2}}\:y_{2}^{m+1} \nonumber \\&&
+\:y_{1}^{m+\frac{5}{2}}\:y_{2}^{m}\:+\:y_{1}^{m+\frac{1}{2}}\:y_{2}^{m+2}\:\big)
  \sqrt{y_{1}+y_{2}}   p(y_{1})\:p(y_{2})\nonumber\\&&
+ \: \frac{1024 \: \sqrt{2}}{15}    +    \frac{8 \: \pi}{3
  \:\tilde{p}_{m+1}^2}     \int^\infty_0 \!\! dy_{1}
   \int^\infty_0 \!\! dy_{2}   
 \big(\ y_{1}^{m+3}\:y_{2}^{m  } \nonumber\\&&
+\:3\:y_{1}^{m+2}\:y_{2}^{m+1}\:\big)
\ln(y_{1}+y_{2})  p(y_{1})\:p(y_{2})\nonumber\\&&
-\frac{8 \: \pi \: \tilde{p}_{m}}{3\:\tilde{p}_{m+1}^2}     \int^\infty_0 \!\! dy   
y^{m+3} \:\ln(y)\:p(y)\nonumber \\&&
-\frac{8 \: \pi }{\tilde{p}_{m+1}}     \int^\infty_0 \!\! dy \;
y^{m+2} \:\ln(y)\:p(y) -\frac{32}{3}\:\ln(2) \nonumber \\&&
+\frac{1}{\tilde{p}_{m+1}^2} \: \int^\infty_0 \!\! dy_{1}
  \int^\infty_0 \!\! dy_{2}  
 \big(\:10\: y_{1}^{m+2}\:y_{2}^{m+1}+\frac{65}{6}
 y_{1}^{m+3}\:y_{2}^{m}\big) \cdot \nonumber\\&& \cdot  \arctan \left(\frac{3}{2}\:\left(\sqrt{\frac{y_{1}}{y_{2}}}+\sqrt{\frac{y_{2}}{y_{1}}}\right)^{-1}\right) p(y_{1})\:p(y_{2})\nonumber\\&&
+\frac{1}{\tilde{p}_{m+1}^2} \: \int^\infty_0 \!\! dy_{1}
 \;\int^\infty_0 \!\! dy_{2} \;
 \left[\big(\:6\: y_{1}^{m+2}\:y_{2}^{m+1}+\frac{21}{2}
 y_{1}^{m+3}\:y_{2}^{m}\big)\cdot \right. \nonumber\\
&&\;\cdot\;\left. \arctan \left(\frac{2}{5}\:\left(\sqrt{\frac{y_{1}}{y_{2}}}-\sqrt{\frac{y_{2}}{y_{1}}}\right)\right)\;p(y_{1})\:p(y_{2}) \right]
-\frac{125}{6} \:\arctan\left(\frac{3}{4}\right)\bigg\}\bigg]\nonumber \\
\end{eqnarray}
The radius of gyration for a polydisperse solution can be obtained from the small momentum behavior of the density correlation 
function \cite{S00}.  This leads to the following average of the radius of gyration $R_g^2(n)$ of an isolated chain  
\begin{eqnarray}
R_g^2[p]&=&\int_0^\infty dy\,\frac{p(y)}{\tilde{p}_2}\,y^2  R^2(yN)\nonumber\\
&=&l_R^2N_R\frac{\tilde{p}_3}{\tilde{p}_2}
\bigg(1+\left(\frac{a_1\sqrt{N_R}
      }{2}-1\right)u+\left(\frac{a_2N_R}{4}+\frac{5}{4}a_1\sqrt{N_R}-\frac{7}{8} \right)u^2 \nonumber \\
&&+\left(\frac{a_3N_R^{\frac{3}{2}}}{8}+\frac{3}{2}a_2N_R+\frac{61}{32}a_1\sqrt{N_R}-\frac{61}{48}
  \right)u^3 \nonumber\\
&&+\left(\frac{a_4N_R^2}{16}+\frac{19}{16}a_3N_R^{\frac{3}{2}}+\frac{83}{16}a_2\,N_R \right. \nonumber\\ 
&&\left. +\left(Z_u(3)+Z_n^{\frac{1}{2}}(3)-\frac{967}{48}\right)\frac{\sqrt{N_R}}{2}a_1+Z_n(4) \right)u^4 +c_g\bigg)\;;\nonumber \\
\end{eqnarray}
\begin{eqnarray}
c_g&=&-\frac{a_1\sqrt{N_R}}{2}\left(1-\int_0^\infty dy
  \frac{p(y)}{\tilde{p}_3}y^{\frac{7}{2}}\right)u\nonumber\\
&&-\left(\frac{a_2N_R}{4}
\left(1-\frac{\tilde{p}_4}{\tilde{p}_3}\right)+\frac{5}{4}a_1\sqrt{N_R}\left(1-\int_0^\infty
  \frac{p(y)}{\tilde{p}_3}y^{\frac{7}{2}}\right) \right)u^2\nonumber\\
&&-\left(\frac{a_3N_R^{\frac{3}{2}} }{8}\left(1-\int_0^\infty dy
    \frac{p(y)}{\tilde{p}_3}y^{\frac{7}{2}} \right) + \frac{3}{2}a_2N_R
  \left(1-\frac{\tilde{p}_4}{\tilde{p}_3} \right) \right. \nonumber\\
&& \left. +\frac{61}{32}a_1\sqrt{N_R}
   \left(1-\int_0^\infty dy \frac{p(y)}{\tilde{p}_3}y^{\frac{7}{2}}\right) \right)u^3-\left(\frac{a_4N_R^2}{16} \left(1-\frac{\tilde{p}_5}{\tilde{p}_3}
  \right)\right. \nonumber\\
 &&+\frac{19}{16}a_3N_R^{\frac{3}{2}}\left(1-\int_0^\infty
    \frac{p(y)}{\tilde{p}_3}y^{\frac{7}{2}}\right)
  +\frac{83}{16}a_2N_R\left(1-\frac{\tilde{p}_4}{\tilde{p}_3}\right)\nonumber\\
&&  \left. +\left(Z_u(3)+Z_n^{\frac{1}{2}}(3)-\frac{967}{48}\right)
  \frac{a_1\sqrt{N_R}}{2} \left(1-\int_0^\infty dy \frac{p(y)}{\tilde{p}_3}y^{\frac{7}{2}}\right)\right)u^4\;,\nonumber \\
\end{eqnarray}
where the coefficients $a_k, Z_u(k)$ and $Z_n^{\frac{1}{2}}(k)$ are taken from (\ref{a2}), (\ref{a5}) and  (\ref{a6}) respectively.
Again the polydispersity correction $c_g$ vanishes for a monodisperse ensemble.
\end{appendix}

\addcontentsline{toc}{chapter}{Literaturverzeichnis}

\bibliographystyle{unsrt}
\bibliography{lit}

\end{document}